\documentclass{article}
\usepackage{emulateapj}
\usepackage{graphicx}
\usepackage{times}
\usepackage{natbib}

\def\kpc{{\rm\thinspace kpc}}

\def\yr{{\rm\thinspace yr}}
\def\kmps{\hbox{$\km\s^{-1}\,$}}
\def\Mpc{{\rm\thinspace Mpc}}

\def\keV{{\rm\thinspace keV}}
\def\kmpspMpc{\hbox{$\kmps\Mpc^{-1}$}}
\def\km{{\rm\thinspace km}}
\def\s{{\rm\thinspace s}}
\def\erg{{\rm\thinspace erg}}
\def\ergps{\hbox{$\erg\s^{-1}\,$}}

\begin{document}

\title{{\em Chandra} ACIS-S observations of Abell 4059: signs of dramatic
interaction between a radio galaxy and a galaxy cluster.}

\author{Sebastian Heinz\altaffilmark{1},
Yun-Young Choi\altaffilmark{2,3,4},
Christopher~S.~Reynolds\altaffilmark{4,5},
and Mitchell C.~Begelman\altaffilmark{4,6}
}

\altaffiltext{1}{Max-Planck-Institut f\"{u}r Astrophysik,
Karl-Schwarzschild-Str.~1, 85740 Garching, Germany, {\tt
heinzs@mpa-garching.mpg.de}}

\altaffiltext{2}{Dept.~of Physics and CAIS, Ewha University, Seoul
120-750, Korea}

\altaffiltext{3}{Center for High Energy Physics, Kyungpook National
University, Daegu 702-701, Korea}

\altaffiltext{4}{JILA, Campus Box 440, University of Colorado,
Boulder CO~80309, Hubble Fellow}

\altaffiltext{5}{Dept. of Astronomy, University of Maryland, College Park
MD 20742}

\altaffiltext{6}{Dept.~of Astrophysical and Planetary Sciences,
University of Colorado, Boulder CO~80309}

\begin{abstract}
{We present {\em Chandra} observations of the galaxy cluster A4059.  We
find strong evidence that the FR-I radio galaxy PKS~2354--35 at the center
of A4059 is inflating cavities with radii $\sim 20\kpc$ in the intracluster
medium (ICM), similar to the situation seen in Perseus A and Hydra A.  We
also find evidence for interaction between the ICM and PKS~2354--35 on
small scales in the very center of the cluster.  Arguments are presented
suggesting that this radio galaxy has faded significantly in radio power
(possibly from an FR-II state) over the past $10^8\yr$.}
\end{abstract}


\section{introduction}

Clusters of galaxies are complex dynamical structures and their cores are
subject to an array of interesting physical processes.  Constraints from
imaging X-ray observations suggest that the hot X-ray emitting intracluster
medium (ICM) in the core regions of rich clusters is radiatively cooling on
timescales shorter than the life of the cluster, giving rise to cooling
flows \citep[][and references therein]{fabian:94}.  The central dominant
galaxy present in many clusters often hosts a radio loud active galactic
nucleus (AGN).  It has been suggested \citep[e.g.,][]{binney:95} that
cooling flows and central cluster radio galaxies are intimately related via
complex feedback processes.  It is easy to see how radio galaxy activity
resulting from black hole accretion can be associated with a cooling flow.
However, the impact of a radio galaxy on its environment is much less
clear.

Theoretically, we expect radio jets to inflate cocoons of relativistic
plasma that expand into the surrounding ICM \citep[e.g.][, hereafter
RHB]{begelman:89,kaiser:97,reynolds:01}.  The energy input by this process
has recently come under investigation for its potential role in heating
cluster cores \citep{bruggen:01,quilis:01,reynolds:01b}.  However, while
our simulations suggest that about half of the energy injected by the jets
can be thermalized in the cluster center, numerical simulations of this
process still carry a large degree of uncertainty, since limited
computational resources require significant simplications.  In order to
verify the validity of the assumptions and to design future models, we
require guidance from observations of radio-galaxy/cluster interactions.

Imaging X-ray observatories, such as the {\it Chandra X-ray Observatory}
(CXO), provide a direct probe of this interaction.  Both {\it ROSAT} and
CXO observations of Perseus~A have found X-ray cavities coincident with the
radio lobes \citep{boehringer:93,fabian:00}, surrounded by X-ray shells
which appear to be slightly cooler than the unperturbed ICM (see, for
example, RHB for a possible explanation).  Similar features are seen in CXO
observations of Hydra~A \citep{mcnamara:00,david:01} and Abell 2052
\citep{blanton:01}.

In this {\it Letter}, we present CXO observations of the rich galaxy
cluster Abell~4059 ($z=0.049$).  The cD galaxy of A4059 hosts the FR-I
radio galaxy PKS~2354--35.  A short {\it ROSAT} High Resolution Imager
(HRI) observation of this source suggested the presence of two ICM cavities
at the same position angle as the radio lobes \citep[][, hereafter
HS]{huang:98}.  In \S~2 we discuss our observations, confirming the
presence of these cavities, and show that A4059 displays significant
additional morphological complexity.  Constraints on models for this source
are discussed in \S~3, \S~4 presents our conclusions.  We assume a Hubble
constant of $H_0=65 \kmpspMpc$ and $q_0 = 1/2$, giving a linear scale of
$1\, {\rm kpc\,arcsec^{-1}} = 0.492\, {\rm kpc\,pixel^{-1}}$.

%
%
\section{Observations}

\subsection{Data Reduction}

The galaxy cluster Abell~4059 was observed with the {\it Advanced CCD
Imaging Spectrometer} (ACIS) on 24-Sept-2000 (22.3\,ksec exposure) and on
4-Jan-2001 (18.4\,ksec exposure).  The radio nucleus of PKS 2354-35 was
placed 1\,arcmin from the nominal aim point of the back-illuminated S3
chip, placing the bulk of the emission on chip-S3.  In this {\it Letter},
we only use data collected on this chip.  The data was read out at the
standard 3.2 sec frame rate, telemetered using faint mode, and filtered on
ASCA event grades. The energy range was restricted to the 0.3-10 keV band
and corrected for exposure and vignetting, spectral fitting was restricted
to 0.3-8.0 keV.

Some of the Sept-2000 data were affected by a period of relatively high
background, with up to 10 counts per second per chip for the
back-illuminated chips.  However, most of the cluster emission is well
above this rate and we decided to use the entire data set for imaging.  For
the spectral analysis, we filtered the data on the counts from chip-S1 to
fall within a factor of 1.2 of the quiescent background rate, which
rejected 33\% of the observing window.  We then used the quiescent
background files by Markevitch\footnote{See {\tt {\footnotesize
http://asc.harvard.edu/cal/Links/Acis/acis/Cal\_prods/bkgrnd/current}}} for
background subtraction.  The background subtracted total count rate on
chip-S3 is 5.1 counts/second.

\subsection{Morphological Appearance}
\label{sec:morphology}
Figure~1a shows an adaptively smoothed image of the central 3 arcmin of
A4059.  Overlayed is the 8\,GHz image by \citep[][, hereafter
T94]{taylor:94}.  While the large scale emission appears smooth (probably
due to the larger smoothing length used by the {\tt csmooth} algorithm),
the core of A4059 is not relaxed: it is double peaked with one peak at the
cluster center and the other 15\,arcsec south-west of the center.  The
smoothed image reveals a distorted hour-glass like structure (yellow
regions in Fig.~1a), centered on the nucleus and oriented perpendicular to
the radio axis.
\begin{figure*}[tbp]
\begin{center}
\resizebox{0.49\textwidth}{!}{\includegraphics{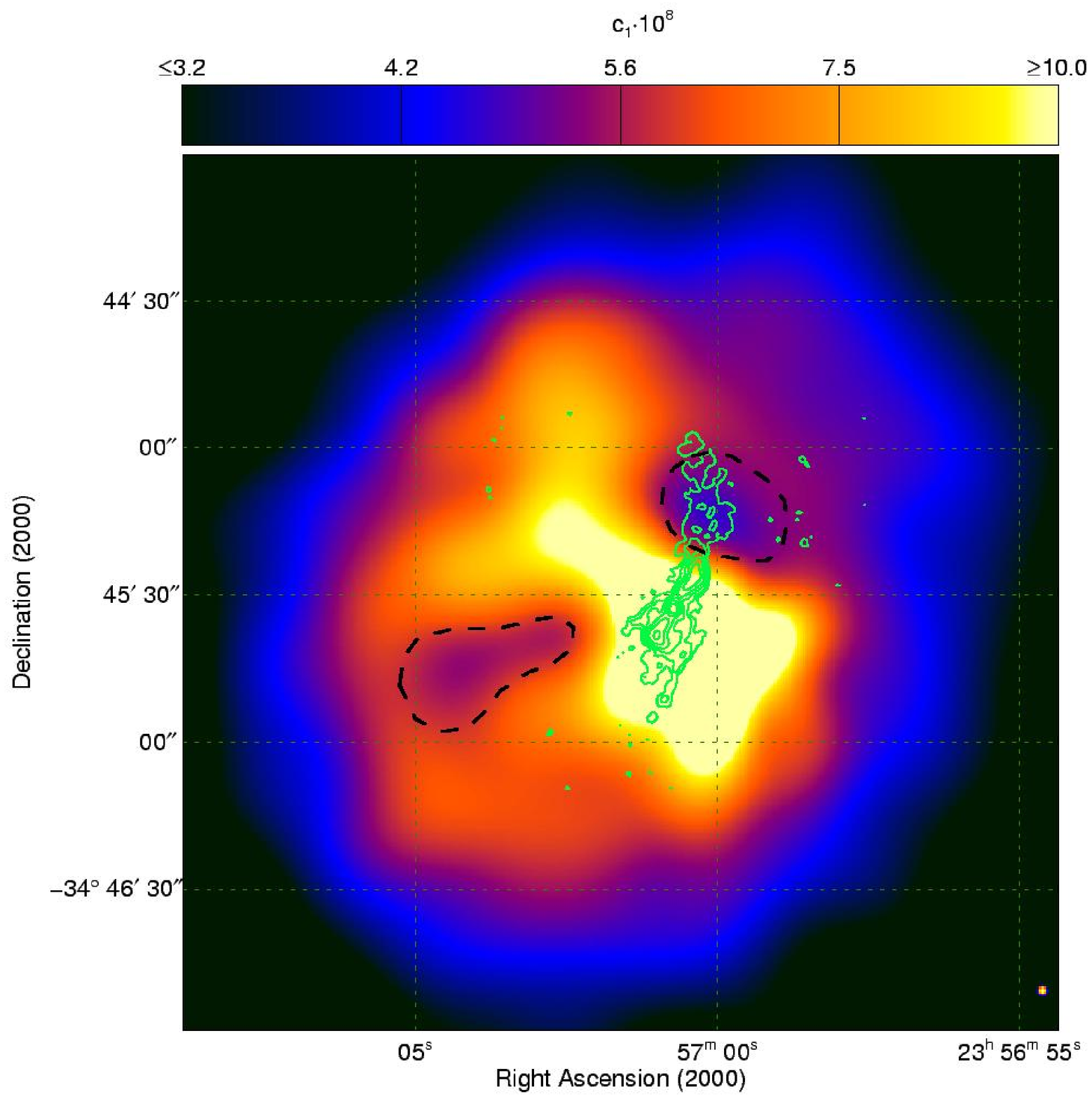}}
\resizebox{0.49\textwidth}{!}{\includegraphics{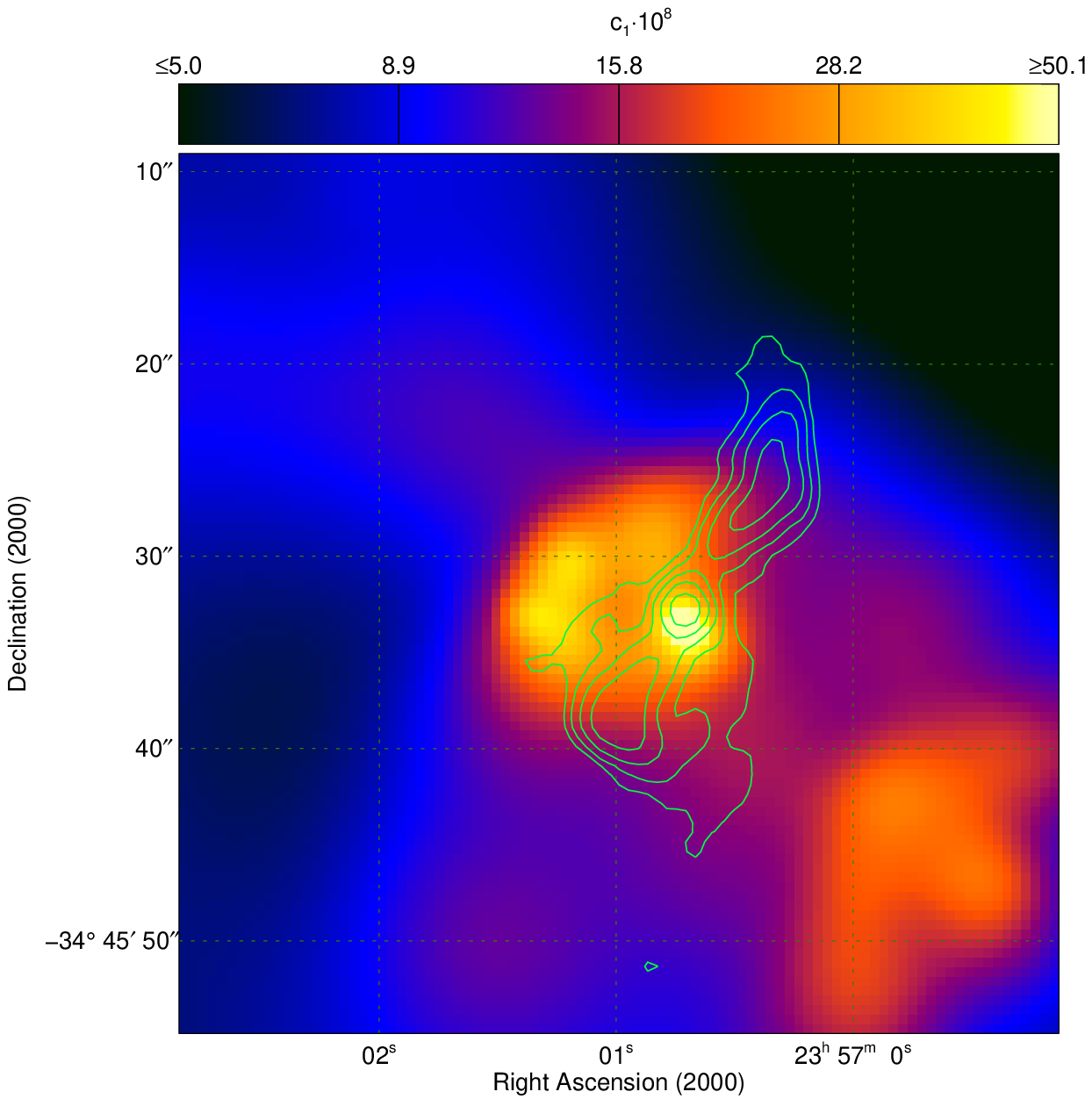}}
\caption{ACIS-S image of A4059. {\em Left panel:} adaptively smoothed
image of $c_{1}$ in units of $10^{-8}\ {\rm cts\ cm^{-2}\ pixel^{-1}\
s^{-1}}$ (significance $5-6\sigma$, $0.492\arcsec$ per pixel) with contours
of 8\,GHz flux (from T94) and contours for statistical evaluation of
cavities (\S\ref{sec:morphology}). {\em Right panel:} adaptively smoothed
representation of the central region (smoothed to $3-4\sigma$ significance,
same units) and 8 GHz radio contours.}
\end{center}
\end{figure*}

The most interesting detail of the X-ray image are the two X-ray holes
already noticed by HS, clearly visible in the 5-sigma smoothed image.  The
higher quality CXO image confirms that these are real local brightness
minima, not just a visual effect caused by a bright central bar
perpendicular to the cluster elongation.  A noticeable offset exists
between the nucleus and the axis between the cavity centers, with the
nucleus being shifted to the south-west by $\sim 12\arcsec$ from this axis.
The cavity centers are roughly $50\arcsec$ apart.

To assess the significance of the cavities we extracted an azimuthally
averaged radial surface brightness profile of the central region, excluding
the cavity regions shown in Fig.~1a (a by-eye approximation) and calculated
the expected surface brightness in the two cavities.  Based on this
estimate, the NW and SE cavities are significant to 26 and 5.3 sigma,
respectively.  However, because the central region of A4059 is strongly
perturbed (roughly inward of 1 arcmin from the center), it is difficult to
make a rigorous statement concerning the significance of the cavities this
way.

However, the outer regions of the cluster appear relaxed.  As an
alternative method, we took radial surface brightness profiles from the
combined SE- and NW quadrants and the combined SW-NE quadrants.  As a first
order correction for cluster ellipticity we shifted the radial SW-NE
profile outward by a factor of 1.13, producing a good match at large radii
with the SE-NW profile.  Because the temperature at large $r$ is relatively
uniform, we used an isothermal $\beta$-model to fit the surface brightness,
which represents the outer cluster very well.  Because the inner cluster
regions are not well represented by a $\beta$-model, we only used points
further than 60 arcsec from the center for the fit ($r_{\rm c} \sim
50\arcsec$, $\beta=0.52$, $\chi^{2}/dof=59/43=1.4$).

We then compared the flux measured inside the cavity contours in Fig.~1a to
that expected from the $\beta$-model.  The NW cavity has 33 sigma
significance compared to the best fit $\beta$-model, while the less
pronounced SE cavity has 13 sigma significance.  Forcing the $\beta$-model
normalization to be consistent with the observed cavity flux increases the
reduced chisquare of the fit by a factor of 2.7 (NW cavity) and 1.9 (SE
cavity), corresponding to a significance of 8.5 and 6 respectively.

The radio overlay in Fig.~1a shows that the 8GHz radio lobes are only
partly coincident with the cavities.  The NW lobe covers a good fraction of
the NW cavity.  The SE lobe is much smaller than the northern lobe and not
spatially coincident with the SE cavity.  We note, however, that
low-frequency radio observations sensitive to spatial scales of
$0.1-1$\,arcmin, which might reveal the full extent of the radio lobes, do
not yet exist.

The central peak contains interesting sub-structure (Fig.~1b., smoothed to
a S/N of 3-4).  The brightest sub peak, which is well resolved by CXO and
has a diameter of 3--4\arcsec $\sim$ 3--4\,kpc, is coincident with the core
of PKS~2354--35 (within the CXO pointing accuracy) and could be emission
from the hot interstellar medium of the central galaxy.  The three
sub-peaks are located around a local brightness minimum.  Our hardness
ratio analysis (\S\ref{sec:spec}) suggests that this minimum is due to lack
of emission rather than intervening absorption.  The close correspondence
between these peaks and the SE radio lobe suggests that this substructure
might be caused by on-going interaction.

\subsection{Spectroscopic Properties}
\label{sec:spec}
We used the adaptive binning routine of \citet{sanders:01} to produce
hardness ratio maps.  We define the hardness ratios $h_{1} = (1-2{\rm \,
keV})/(0.3-1{\rm \, keV})$ and $h_{2} = (2-10{\rm \, keV})/(1-2{\rm \,
keV})$ as the ratio of the counts in the respective energy bands and
$c_{1}$ as the surface brightness in the 0.3-10 keV band in ${\rm cts\
cm^{-2}\ s^{-1}\ pixel^{-1}}$.  For ICM observations, $h_1$ is mainly
sensitive to absorption variations, whereas $h_2$ is a temperature
diagnostic.  While there is no sign of varying absorption across the field
(as seen in $h_1$), the $h_2$ map clearly shows a radial temperature
gradient, as would be expected in a cooling flow cluster (Fig.~2).  It also
shows a global temperature gradient, with the SW half of the image
appearing hotter than the NE half.

We have extracted a global spectrum of A4059 from the central 90\,arcsec.
Fitting this spectrum with a two temperature {\tt wabs*zwabs*(mekal+mekal)}
thermal plasma model (Galactic neutral hydrogen column fixed to $N_{\rm
H,G} = 1.45\times 10^{20}\,{\rm cm^{-2}}$) results in the best fit
parameters $kT_1=1.34^{+0.53}_{-0.19} \keV$, $kT_2=3.90^{+1.19}_{-0.36}
\keV$, $Z=0.60^{+0.16}_{-0.11}$, $N_{\rm H,z}=5.36^{+0.48}_{-0.54}\times
10^{20}\,{\rm cm^{-2}}$, $\chi^2/dof=1.29$ (3 sigma error bars). A {\tt
wabs*zwabs*(mkcflow + mekal)} cooling flow model provides a similarly
reasonable fit ($kT_{\rm 1} = 0.1^{+0.75}_{-0.1}\,{\rm keV}$, $kT_{\rm
2}=kT_{\rm {\tt mekal}}=3.80^{+0.16}_{-0.13}\,{\rm keV}$,
$Z=0.71^{+0.09}_{-0.09}$, $\dot{M} = 27.6^{+6.0}_{-5.9}\, M_{\sun}{\rm \,
yr^{-1}}$, $N_{\rm H,z}=5.54_{-0.40}^{+0.25} \times 10^{20}\,{\rm
cm^{-2}}$, $\chi^2/dof = 1.26$).

\section{Discussion}

Like Perseus~A and Hydra~A, A4059 shows clear signs of interaction between
the radio galaxy and the ICM.  However, unlike in these systems, there is
no one-to-one correspondence between the radio lobes and the X-ray holes.
We argue below that the cavities were indeed created by the radio activity
(given the similarity in position angle and the fact that there are no
other viable models for producing large ICM cavities), but that the radio
galaxy has faded since they were formed.  This is particularly interesting
since CXO has recently found clusters with cavities but without a radio
source \citep[e.g.,][]{mcnamara:01}: A4059 could be a `missing link'
between cavities with and without detectable radio lobes, supporting the
notion that cluster cavities can be created by radio galaxies without
having to show current radio activity.

\subsection{Evolutionary state}

In the hydrodynamic simulations of RHB, we identified three evolutionary
phases.  Early on, the cocoon is highly overpressured, driving a strong
shock into the ICM.  Simulated X-ray maps show a cavity surrounded by a
thin, hot shell.  Once the cocoon comes into pressure equilibrium with the
ICM, the sideways expansion of the cocoon becomes sub-sonic and the shock
becomes a compression wave, though active jets can keep the advance speed
of the jet heads supersonic for much longer, driving ``sonic booms'' into
the ICM.  For roughly a sound crossing time, the well defined cavity
created during the supersonic phase will survive and be observable.  Then,
hydrodynamic instabilities will destroy the cavity altogether.

A strong shock would show up as a sharp feature in our X-ray images, which
have an effective resolution\footnote{After adaptive smoothing to a signal
to noise of 5.} of 4--5\,arcsec at the edges of the cavities.  The absence
of such a shock indicates that the source expansion is no longer highly
supersonic.  We identify the hourglass-like feature in the core with the
compression wave (``sonic boom'') found in our hydrodynamic simulations
(Fig.~1, RHB; in our simulated X-ray maps the brightest emission also tends
to be in the equatorial plane).

\subsection{Source Power}

We can estimate the radio source parameters based on the presence of the
X-ray cavities (\citealt{heinz:98}, RHB, \citealt{churazov:00}).  We use
the two-temperature fit of \S~2 to estimate the physical parameters of the
ICM.  Taking the hot emission to arise uniformly in a sphere of 90\arcsec\
radius yields an electron density of $n_{\rm hot}\gtrsim 0.009\,{\rm
cm^{-3}}$.  Assuming the cold gas is in pressure equilibrium with the hot
gas gives an electron density of $n_{\rm cold} \gtrsim 0.031\, {\rm
cm^{-3}}$ and a volume filling factor of $5\times 10^{-3}$.
\begin{figure*}[tbp]
\begin{center}
\resizebox{0.5\textwidth}{!}{\includegraphics{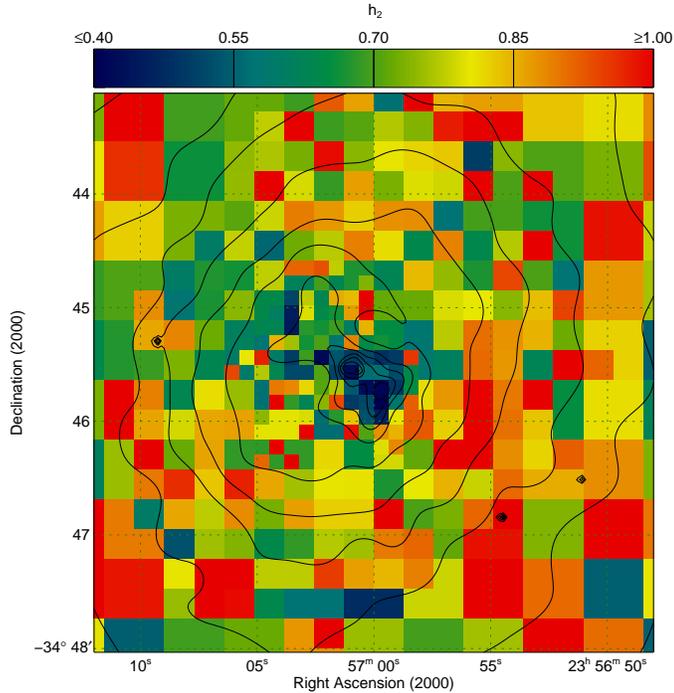}}
\caption{Hardness ratio $h_2$ adaptively binned to reach a signal-to-noise
of 5 (after background subtraction) with contours from the $5-6\sigma$
adaptively smoothed X-ray image (contour start from $c_{1}=10^{-8}\,{\rm
ergs\,cm^{-2}\,s^{-2}\,pixel^{-1}}$, increasing by a factor of $1.4$ between
each contour).}
\end{center}
\end{figure*}

We assume that both X-ray cavities are completely evacuated by the lobes
and estimate their size from the smoothed images by approximating them as
spheres.  While this is clearly a simplification, it will be sufficient for
this order of magnitude estimate.  A `by eye' fit of the cavities gives
bubble radii of $r_{\rm bub} \sim 20\arcsec$ (20 kpc).  The pressure in the
hot phase is $p_{\rm ICM} \gtrsim 1.1\times 10^{-10}\,{\rm erg\,cm^{-3}}$,
relatively close to the minimum energy pressure in the lobes of $p_{\rm ME}
\sim 3 - 5\times 10^{-11}{\rm ergs\ cm^{-3}}$ (T94).  At a minimum, the
radio galaxy has to perform ``p\,dV'' work against the ICM.  Including the
internal energy of the plasma within the cavities, this gives an integrated
energy output of $E_{\rm tot} \gtrsim 8\times 10^{59}\, {\rm ergs}$.

The radio galaxy had to inject this energy into the bubbles before they
floated out of the cluster core.  This buoyancy timescale is approximately
twice the sound crossing time of the relevant region of the cluster,
$\tau\sim 4r_{\rm bub}/c_{\rm s}\sim 8\times 10^{7}\yr$, where we have used
the sound speed for a $4\keV$ gas, $c_{\rm s}\sim 1000\kmps$.  The
time-averaged source power needed to produce the cavities is then $L_{\rm
kin}\sim E_{\rm b}/\tau\gtrsim 3\times 10^{44}\ergps$.

Alternatively, we can estimate the source age from the ``sonic boom''
arguments of RHB.  The hour-glass structure through the cluster center is
roughly 50\arcsec\ long (i.e., $25\arcsec \sim 25\,{\rm kpc}$ on either
side of the center).  Following RHB, we equate this to the distance
traveled by a shock/compression wave which moves at least at the sound
speed of the hot ICM.  This gives a source age of $\tau \lesssim 2.4\times
10^{7}\,{\rm yrs}$ and a time-averaged power of $L_{\rm kin}\gtrsim E_{\rm
b}/\tau\sim 10^{45}\ergps$.

A third, X-ray independent way to estimate the source power is based on the
radio flux.  The flux densities at 5\,GHz and 8\,GHz are given by T94 as
76\,mJy and 34\,mJy, while the 1.4\,GHz NVSS flux \citet{condon:98} is
1.3\,Jy.  The NVSS flux lies a factor of 2 above the extrapolation of the
5--8\,GHz flux, and the NVSS image suggests spatial extension on arcmin
scales (a factor of $\sim 2$ larger than seen at 5--8\,GHz).  This suggests
that NVSS is detecting low frequency emission from plasma that is emitting
a steep radio spectrum, possibly indicating that it has suffered
synchrotron aging.  A reasonable upper limit on the current radio power can
be derived by taking the 1.4\,GHz luminosity, and using the arguments of
\citet{bicknell:98} to convert it into a kinetic luminosity.  Taking the
smallest reasonable value of their conversion parameter, $\kappa_{1.4} >
10^{-12}$, we estimate an upper limit on the current kinetic power of
$L_{\rm kin} < 7\times 10^{43}\ergps$.

Comparing the time-averaged source power (derived the X-ray cavities) to
that derived from the radio luminosity (which is equivalent to the source
power averaged over the synchrotron cooling time of the 1.4\,GHz electrons,
$\tau_{\rm cool} \lesssim 10^7\,{\rm yrs}$), one infers that either this
source has faded in kinetic luminosity by an order of magnitude or more, or
that the magnetic field in the lobes is considerably out of equipartition.
Since the thermal pressure is close to the equipartition pressure estimated
by T94, we favor the first possibility.  Given the uncertainties in these
arguments (especially in $\kappa_{1.4}$, for which we chose a conservative
value), the source could easily have faded by more than an order of
magnitude.  Indeed, the fact that the average power is in the realm of
FR-II radio galaxies, while morphology and current radio luminosity qualify
it as an FR-I, leads us to speculate that PKS2354--35 is an example of an
FR-II source that has faded into an FR-I source on a timescale of less than
$10^8\yr$.

The apparent offset between the cluster center and the center of the
cavities and the asymmetric brightness distribution through the equatorial
regions of the hour-glass structure may be evidence for bulk ICM motions.
In particular, the morphology suggests a bulk flow in a NE direction which
might further squeeze the outward moving compression wave from the radio
galaxy, and sweep back the cavity structure.  We note that the SW ridge is
rather cool and thus cannot be a strong shock resulting from the
interaction of a bulk flow with the radio galaxy.  Hydrodynamic simulations
are required to investigate this system further.

\section{Conclusions}

We have presented CXO observations of Abell~4059.  While the ICM appears
smooth and relaxed on large scales, it shows complex morphology in the core
region which is likely the result of interaction between the ICM and the
central FR-I radio galaxy PKS2354--35.  As was suggested by HS,
PKS~2354--35 appears to have inflated two large cavities in the ICM.
Together with a central bar-like structure, these cavities produce an
hour-glass like morphology which can be readily understood as being due to
a radio cocoon expanding into the ICM.  While clear correspondence exists
between the NW cavity and the NW radio lobe, the SE cavity is much larger
than the SE lobe, suggesting that this could be a `missing link' between
cavities with and without visible radio lobes.

The absence of sharp edges in the brightness images and of large
temperature jumps implies that PKS2354--35 is {\it not} driving a strong
shock into the ICM.  We suggest that it is in the
weak-shock/compression-wave phase identified in the hydrodynamic
simulations of RHB.  Dynamical estimates give a time averaged kinetic
source power of at least $L_{\rm kin} \gtrsim 3\times 10^{44}\ergps$, while
estimates based on the current radio luminosity indicate a source power of
$L_{\rm kin} \lesssim 7\times 10^{43}\ergps$.  We suggest that this source
has faded by a significant amount (and possibly from an FR-II phase) during
the past $10^8\yr$.

\section*{Acknowledgments}

We thank Eugene Churazov and Torsten Ensslin for helpful discussions.  We
acknowledge support from SAO grant GO0-1129X, the National Science
Foundation grants AST~9529170 and AST~9876887, and NASA under grant
NAG5-6337.  YC wishes to thank support from the MOST through the National R
\& D program for women's universities.  CSR thanks support from Hubble
Fellowship Grant HF~01113.01-98A.

\end{document}